\documentclass{eas}
\usepackage{graphicx}
\begin{document}
\title{The Gaia-LSST Synergy} 
\author{\v{Z}eljko Ivezi\'{c}}\address{LSST Project Scientist, University of Washington, Seattle, WA 98155; LSST Project Office, 933 N. Cherry Avenue, Tucson, AZ  85721}
\author{Steven M. Kahn}\address{LSST Director, Kavli Institute for Particle Astrophysics and Cosmology, SLAC National
                          Accelerator Laboratory, Stanford University, Stanford, CA 94025; LSST Project Office, 933 N. Cherry Avenue, Tucson, AZ  85721}
\author{Patricia Eliason}\address{Executive Officer for LSST Corporation, 933 N. Cherry Avenue, Tucson, AZ 85721} 
\begin{abstract}
We discuss the synergy of Gaia and the Large Synoptic Survey Telescope (LSST) in the context of
Milky Way studies. LSST can be thought of as Gaia's deep complement because the two surveys
will deliver trigonometric parallax, proper-motion, and photometric measurements with similar
uncertainties at Gaia's faint end at $r=20$, and LSST will extend these measurements to a limit about
five magnitudes fainter. We also point out that users of Gaia data will have developed data
analysis skills required to benefit from LSST data, and provide detailed information about how 
international participants can join LSST. 
\end{abstract}
\maketitle

\section{Introduction}

Studies of stellar populations, understood to mean collections of stars with common spatial, kinematic, 
chemical, and/or age distributions, have been reinvigorated during the past decade by the advent of 
large-area sky surveys such as the Sloan Digital Sky Survey and the Two-Micron All Sky Survey (for a 
recent review, see Ivezi\'{c}, Beers \& Juri\'{c} 2012, hereafter IBJ2012). The quest for more precise and 
time-resolved data over large sky areas continues with the advent of Gaia and LSST. 

Gaia is an ESA Cornerstone space mission that will survey the sky to a magnitude limit of $r\sim20$ and obtain 
astrometric and three-band photometric measurements for about 1 billion sources, as well as radial-velocity 
and chemical-composition measurements for about 100 million brightest stars (Perryman et al. 2001). 
The LSST will be a wide-field ground-based optical imaging system with an 8.4m primary mirror, 
a 9.6 deg$^2$ field of view, and a 3.2 Gigapixel camera. LSST will deliver a six-band survey of about 18,000 deg$^2$
of the sky visible from Cerro Pachon in Chile to a depth of $r=27.5$. Over 800 observations of each position 
(summed over all six bands) that will be obtained during the 10-year survey will enable unprecedented 
time-domain studies of the faint optical sky. This massive dataset will include photometric and astrometric data 
for about 20 billion Milky Way stars. For a more detailed discussion, including optical design, the filter complement, 
the focal plane layout, and special science programs, please see the LSST overview paper (Ivezi\'{c} et al. 2008) and 
the LSST Science Book (Abell et al. 2009).

Within the context of Milky Way studies, Gaia and LSST datasets will be highly complementary and synergistic,
as discussed in the next section. In addition to scientific connections, the data analysis skills developed by 
users of Gaia data will carry over to the LSST era -- it is likely that Gaia users will be among the most efficient and
productive users of LSST data. We discuss possibilities for international participation in LSST in the
following two sections and summarize our conclusions in the last section.

\section{The Gaia-LSST Synergy}

In the context of Gaia, the LSST can be thought of as its deep complement (for a detailed discussion, we refer
the reader to IBJ2012). Gaia will provide an all-sky catalog with unsurpassed trigonometric parallax, proper-motion, 
and photometric measurements to r$\sim$20, for about 1 billion stars. The LSST will extend the static map to r$\sim$27 
over half of the sky, detecting about 20 billion stars. A detailed comparison of LSST and Gaia measurement uncertainties 
is given in Figure 1. The LSST will obtain photometric, proper-motion and trigonometric parallax measurements of comparable 
accuracy to those of Gaia at Gaia's faint limit and smoothly extend Gaia's uncertainty versus magnitude 
curves deeper by about 5 mag. Because of Gaia's superb astrometric and photometric quality and the LSST's significantly 
deeper reach, the two surveys are highly complementary -- Gaia will map the Milky Way's disk with unprecedented detail, 
and the LSST will extend this map all the way to the edge of the known halo and beyond.

For example, stars just below the main-sequence turn-off, with $M_r=4.5$, will be detected by Gaia to a distance 
limit of $\sim$10 kpc ($r<20$), and to $\sim$100 kpc with LSST's single-epoch data ($r<24.5$). For intrinsically
faint stars, such as late M dwarfs, L/T dwarfs, and white dwarfs, the deeper limit of LSST will enable detection and 
characterization of the halo populations (for illustration, see Figure 23 in IBJ2012). A star with $M_r=15$ will 
be detectable to a distance limit of 100 pc with Gaia and $\sim$800 pc with LSST, hence the LSST samples will be
about 100 times larger. In addition, for a substantial fraction of red stars with $r>20$, LSST will provide trigonometric 
parallax measurements accurate to better than 10\%. Therefore, despite the unprecedented performance of Gaia for 
$r<20$, LSST will enable major discoveries with its deep $r>20$ sky coverage. At the same time, and in addition to 
its own discoveries, Gaia will provide excellent astrometric and photometric calibration samples for LSST.

\begin{figure}%
\includegraphics[width=12cm]{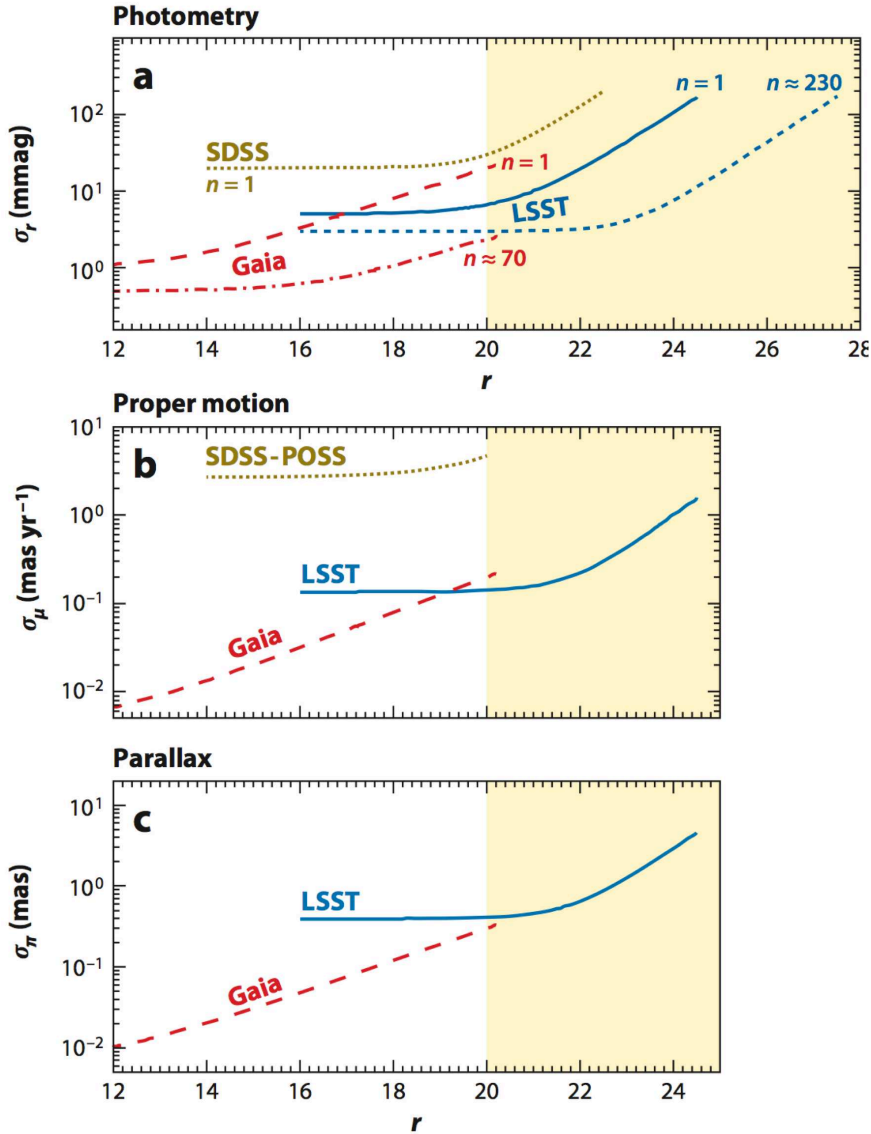}
\vskip -0.2in
\caption{A comparison of the photometric, proper-motion, and trigonometric parallax 
measurement uncertainties for SDSS, Gaia, and LSST, as a function of apparent magnitude 
$r$, for a G2V star. In the top panel, the curve marked ``SDSS'' corresponds
to a single SDSS observation. The red curves correspond to Gaia; the
long-dashed curve shows a single-transit accuracy, while the    
dot-dashed curve shows the end-of-mission accuracy (assuming 70 transits).
The blue curves correspond to LSST; the solid curve shows a single-visit accuracy,
while the short-dashed curve shows the accuracy for co-added data (assuming 
230 visits in the $r$ band). Adapted from \cite{IBJ2012}.}
\label{LSSTvsGaia}
\end{figure}

\section{Opportunities for International Participation in LSST}

LSST was designed to be a public project, with full access to all data and data products\footnote{LSST
Data Products are defined in a living document available as http://ls.st/dpdd}
open to the entire U.S. scientific community, as well as the public at large. Because the observatory
will be sited in Chile, the Chilean astronomical community fully participates in the Project, 
with equal status and data access as their colleagues from the U.S. In addition, a limited 
number of French scientists (from laboratories associated with the Institut national de physique
nucl\'{e}aire et de physique des particules —- IN2P3) have been granted 
access in view of French contributions to the development of the LSST camera\footnote{The
LSST construction proposals have been submitted and approved by the two U.S. federal agencies, 
and the appropriate funding profiles have been defined to complete the Project (see 
\S\ref{sec:construction}). The LSST Project is therefore no longer considering “in-kind” contributions 
to the LSST construction.}.

Unlimited immediate access to LSST Data Releases will also be granted to international 
partners who signed Memoranda of Understanding (MOUs) or Memoranda of Agreement (MOAs) with LSST.  
For other users of LSST 
data, there will be a 2-year delay\footnote{The transient stream, based on image difference
analysis, will be available to everyone in the world within 60 seconds from closing the shutter.}. 
Here we briefly overview background on LSST funding and provide detailed information about 
how international participants can sign MOUs and thus secure immediate access to Data Releases 
(and obtain an opportunity to join LSST science collaborations, see below). 

\subsection{LSST Construction Funding \label{sec:construction}} 

The LSST construction funding is primarily coming from two U.S. federal agencies: the National 
Science Foundation (NSF) and the Department of Energy (DOE). The NSF contribution is 
funded under the Major Research Equipment and Facility Construction (MREFC) line (total not 
to exceed cost is \$473M) and it supports the telescope and site facility construction, 
the data management system, and the education and public outreach components.
This is funded through a cooperative agreement with the Association of Universities for Research in Astronomy
(AURA), which also manages the National Optical Astronomy Observatory (NOAO), the 
Space Telescope Science Institute (STScI), and other facilities. The DOE contribution is
funded as a Major Item of Equipment (MIE) through the Office of High Energy Physics
in the Office of Science, and it supports the camera fabrication (total not to exceed cost 
is \$168M). SLAC National Accelerator Laboratory is the lead DOE laboratory for the 
fabrication of LSST camera.

LSST Corporation has also received substantial support from private donors (about \$40M),
which enabled early development of the primary/tertiary mirror, the secondary mirror blank, 
preliminary site preparation, as well as early sensor studies and some data management 
activities. Key donors include the Lisa and Charles Simonyi Fund for Arts and Sciences, 
Bill Gates, Richard Caris, the W.M. Keck Foundation, Research Corporation for Science
Advancement, Wayne Rosing and Dorothy Largay, Eric and Wendy Schmidt, and Edgar Smith.

\subsection{LSST Operations Funding} 

The NSF MREFC and DOE MIE funds only cover project construction and fabrication costs.
The money allocated cannot be used for operations or for the conduct of science with the 
facility. Instead, operations funding comes from the Astronomical Sciences division budget 
at NSF, and from Cosmic Frontier research funding in the Office of High Energy Physics at DOE. In both cases, there
is significant competition with other projects for those resources.

The current estimate for the annual operations costs of LSST is $\sim$\$37M/yr in 2013 US\$. 
This estimate includes mountain and base facility operations, as well as the costs associated 
with the data processing and data access. It does not include costs for ``doing science'' with
LSST data. Those funds will come from the grants programs at the two agencies. 

The expected support for the LSST operations by NSF and DOE, compared to the above estimate,
will fall short by about \$10M/yr (again in 2013 US\$). {\it LSST is currently seeking international 
partners to support this cost differential.}

\subsection{International Support for LSST Operations} 

LSST is seeking international partners to support about a quarter of the operations costs, in the 
amount of \$10M/yr, with expected start of operations in 2021. With an anticipated number of 
interested Principal Investigators (PI) set to 500,  the cost per PI is \$200,000 integrated over 
ten years (plus a much smaller amount to offset data distribution costs, see below). Here ``a single PI'' assumes 
a senior researcher plus two postdocs and two graduate students. This contribution can be paid 
in full or in up to ten yearly installments of \$20K, with the payments starting any time between 
now and 2019. Contributions received prior to the start of operations will be placed in escrow in 
a U.S. bank. If for any reason the project does not proceed into operations as planned, that money 
would then be returned to its original contributors. All of the figures above are in 2013 US\$, and 
must be appropriately escalated to the time of payment.  International partner institutions will sign 
MOUs with the LSST Corporation.

\subsection{The Roles of LSST Corporation} 

LSST Corporation (LSSTC) was formed in 2003 as a non-profit organization to further the development 
of the LSST Concept. With the onset of construction, LSSTC will be primarily concerned with enabling 
LSST science (by hosting workshops, conducting training programs, and supporting student and 
postdoctoral fellowships as prioritized by its member institutions).

Per agreement with the federal funding agencies, LSSTC is also the vehicle through which international
partner institutions will contribute to operations funding. LSSTC is currently 
negotiating MOUs with such organizations to solidify these commitments. Due to the deadline for 
submitting a proposal for operations to the funding agencies, {\it this process needs to be completed 
by early 2016}. 

Once an agreement with LSSTC is signed (or that it is clear that it will be signed), investigators from a 
foreign institution will be welcomed into the LSST community, including full access to LSST websites, 
simulations, software, 
etc. 

In cases of sizable institutional participation in LSST, foreign institutes might also consider applying to join LSSTC 
as a member institution and thus participate in LSSTC governance. Membership will entitle that institution 
to have a representative attend LSSTC Board meetings (monthly telecons and two face-to-face meetings 
per year). Attendance at those meetings gives the members institutions a significant role in the oversight
of the project, and particularly over science planning associated with the project.

LSSTC will have a significant annual budget that can be used during the construction phase ``for the scientific 
optimization'' of LSST. The details are still being worked out, but expected activities include
workshops on selected LSST-related science topics, training programs for students and postdocs to 
familiarize them with LSST software and data analysis tools, fellowships in LSST science, and sponsoring 
scholarly reports on LSST scientific topics.

Application for membership is via a letter explaining the interests of and capabilities of the institute staff in 
LSST science, and a commitment to the initial and annual 
payments (an initial fee of \$75K, and an annual dues payment of \$25K).

\subsection{The Cost of Accessing LSST Data} 

At present, the construction of two LSST data access centers, one at the National Center for Supercomputing
Applications in Illinois and one in La Serena in Chile, is planned. These have been sized to handle the 
expected number (and spectrum) of scientific queries from the U.S. and Chilean communities. With 
user base enlarged with international partners,  there is need to expand these facilities (with costs 
incurred by new partners), or have international partners build their own data access centers to serve their 
own communities. At this time, the approximate cost estimates are as follows: \$1,500 per year per PI for 
U.S.-based data access and about \$3.5M to set up an independent data access center from 
scratch (using software provided by LSST).

\section{The Benefits of Participation in LSST} 

Although the current call for international partners of LSST is motivated by the immediate need to 
close the operations budget, the LSST Project is fundamentally looking for partners, rather than
selling data. LSST will be a world unique scientific facility and the scientific exploitation of LSST data
will certainly benefit from greater international participation. New international collaborators 
will bring creative ideas for innovative investigations with LSST, access to corollary facilities that 
can enhance the science of LSST, and key skills to collaborations that are preparing for some of the 
more challenging LSST analyses. For example, it is highly likely that Gaia users will be among the most 
efficient and productive users of LSST data because they will have developed data analysis skills 
required to handle a billion-star Gaia survey. 

Preparing for LSST science is a major effort over the next few years. Waiting for operations to begin 
before getting engaged will probably not be a successful path to scientific leadership in LSST science.
To facilitate preparation for LSST science, joining one or more of the existing science collaborations is 
probably the best route to engagement as there is no substitute for direct communication between 
scientists, and we expect international collaborators to get engaged early. 

\subsection{LSST Science Collaborations}

A set of LSST science collaborations was formed by the Project in 2006 with the intent to provide a forum
to engage the community in interacting with the project team (e.g. to discuss science opportunities and 
challenges in each of the main science areas that LSST will address). These collaborations played an essential 
part in producing the LSST Science Book in 2009, which was instrumental for LSST achieving the top ranking 
for ground-based astronomical projects by the U.S. Decadal Survey Committee in 2010.

At this stage, the LSST Project no longer takes any oversight role for the collaborations -- they are free to 
set their own policies for admission, governance, publication, etc. The only constraint is that their members 
must have direct access to LSST data. Once an MOU with LSSTC is signed by international partners, 
they will be welcomed into the LSST community, including full access to LSST websites, simulations, 
software, science collaborations, etc. 

We emphasize that there is no requirement against the formation of new collaborations in new areas, or 
even multiple competing collaborations in the same areas. It is left to the community to figure out how 
best to work with one another in preparing for LSST science.

\section{Conclusions} 

Within the context of Milky Way studies, Gaia and LSST datasets will be highly complementary and synergistic.
In addition to scientific connections, it is likely that Gaia users will be among the most efficient and
productive users of LSST data because they will have already developed data analysis skills required to handle 
a billion-star Gaia survey. LSST is currently seeking international partners and we encourage the
interested colleagues to contact the LSST Corporation.


\end{document}